\documentstyle[aps,psfig,epsf,bbox]{revtex} \topmargin=0in \headheight=0in
\begin{document}

\newcommand{\vAi}{{\cal A}_{i_1\cdots i_n}} \newcommand{\vAim}{{\cal
A}_{i_1\cdots i_{n-1}}} \newcommand{\vAbi}{\bar{\cal A}^{i_1\cdots i_n}}
\newcommand{\vAbim}{\bar{\cal A}^{i_1\cdots i_{n-1}}}
\newcommand{\htS}{\hat{S}} \newcommand{\htR}{\hat{R}}
\newcommand{\htI}{\hat{I}}
\newcommand{\htB}{\hat{B}} \newcommand{\htD}{\hat{D}}
\newcommand{\htV}{\hat{V}} \newcommand{\cT}{{\cal T}} \newcommand{\cM}{{\cal
M}} \newcommand{\cMs}{{\cal M}^*} \newcommand{\vk}{{\vec k}}
\newcommand{\vK}{{\vec K}} \newcommand{\vb}{{\vec b}} \newcommand{{\vp}}{{\vec
p}} \newcommand{{\vq}}{{\vec q}} \newcommand{\vQ}{{\vec Q}}
\newcommand{\vx}{{\vec x}}
\newcommand{\tr}{{{\rm Tr}}} \newcommand{\beq}{\begin{equation}}
\newcommand{\eeq}[1]{\label{#1} \end{equation}} \newcommand{\half}{{\textstyle
\frac{1}{2}}} \newcommand{\gton}{\stackrel{>}{\sim}}
\newcommand{\lton}{\mathrel{\lower.9ex \hbox{$\stackrel{\displaystyle
<}{\sim}$}}} \newcommand{\ee}{\end{equation}}
\newcommand{\ben}{\begin{enumerate}} \newcommand{\een}{\end{enumerate}}
\newcommand{\bit}{\begin{itemize}} \newcommand{\eit}{\end{itemize}}
\newcommand{\bc}{\begin{center}} \newcommand{\ec}{\end{center}}
\newcommand{\bea}{\begin{eqnarray}} \newcommand{\eea}{\end{eqnarray}}
\newcommand{\beqar}{\begin{eqnarray}} \newcommand{\eeqar}[1]{\label{#1}
\end{eqnarray}} \newcommand{\bra}[1]{\langle {#1}|}
\newcommand{\ket}[1]{|{#1}\rangle}
\newcommand{\norm}[2]{\langle{#1}|{#2}\rangle}
\newcommand{\brac}[3]{\langle{#1}|{#2}|{#3}\rangle} \newcommand{\hilb}{{\cal
H}} \newcommand{\pleft}{\stackrel{\leftarrow}{\partial}}
\newcommand{\pright}{\stackrel{\rightarrow}{\partial}}

\begin{flushright}
\vskip .5cm
\end{flushright} \vspace{1cm}

\begin{center}
{\Large {REACTION OPERATOR APPROACH TO MULTIPLE \\ \vskip .5cm 
ELASTIC SCATTERINGS}}

\vspace{1cm}

{ M. Gyulassy$^{1}$, P. Levai$^{2}$, and I. Vitev$^{1}$ }

\vspace{.8cm}

{\em {$^1$ Department of Physics, Columbia University, 538 West
 120-th Street,\\ New York,
  NY 10027, USA\\ $^2$ KFKI Research Institute for Particle and Nuclear
  Physics, \\ P.O. Box 49, Budapest 1525, Hungary} } \vspace{.5cm}

\end{center}

\vspace{.5cm}

\begin{abstract}
We apply the GLV 
Reaction Operator formalism to
compute the effects of multiple elastic scatterings of jets propagating 
through dense matter.  We derive the elastic Reaction Operator 
and  demonstrate that the recursion relations have a closed
form solution that reduces to the familiar Glauber form. We also
investigate  the accuracy of the Gaussian dipole approximation for 
jet transverse momentum broadening. 

\vspace{.5cm} {\em PACS numbers:} 12.38.Mh; 24.85.+p; 25.75.-q
\end{abstract}
\vspace{.5cm}

\section{Introduction}

New  experiments at the Relativistic
Heavy Ion Collider (RHIC) will provide  high statistics
high $p_T$ measurements of photon, lepton and hadron production in
$A+A$ reactions 
\cite{Adcox:2001jp,Adler:2001yq}.  These observables
provide novel tests of pQCD multiple scattering theory.
Quantitative comparisons~\cite{Dumitru:2001jx,Vitev:2001zn,Gyulassy:2000gk} 
to data  
require  understanding of nuclear effects such as  parton 
shadowing~\cite{Eskola:1998df}, the Cronin 
effect~\cite{Cronin:zm,Zhang:2001ce,Accardi:2001ih,WangQ} 
and non-abelian jet energy  
loss~\cite{Wang:xy,Baier:1997kr,Gyulassy:2000fs}. 
In this paper  we apply the Reaction Operator method~\cite{Gyulassy:2000fs},
 developed to solve the radiative energy loss problem, to derive
 the well known multiple elastic Glauber  distribution~\cite{Glauber:1970jm}
in order to illustrate the power of this
recursion technique. The summation
to all orders in the opacity,
$\chi = \int \sigma(z) \rho(z) dz = L/\lambda$, of the matter is 
given in closed form.
In addition we evaluate the accuracy of the  dipole
approximation, leading to the Gaussian approximation to Moliere scattering,
for the case of moderate opacities $\chi<10$ relevant
for applications in nuclear physics.

We describe multiple jet scattering as a series expansion in 
$\chi $,  the  mean number 
of scatterings that a hard parton undergoes along its 
trajectory~\cite{Gyulassy:2000fs,Wiedemann:2000za}.  Jet
production is modeled here by an initial 
 wave packet $j(p)$ for a spinless parton
in color representation $R$ prepared at finite time $t_0$ and centered
at $\vec{\bf x}_0 = (z_0,{\bf x}_0)$.
The momentum space amplitude 
in the absence of 
interactions is then given by
\beq
M_0\equiv i  e^{ipx_0} j(p) \, {\bf 1}_{d_R \times d_R}\;\;,
\eeq{p0}
where the $d_R$ dimensional unit matrix accounts for the jet  colors.
Multiplying $|M_0|^2$ by the invariant one particle phase space
element $d^3\vec{\bf p}/((2\pi)^3 2|\vec{\bf p}|)$ and taking the 
color trace one arrives at the unperturbed 
inclusive distribution of jets in the wave packet
\beq
d^3N_0= {\rm Tr}\;
|M_0|^2 \frac{d^3 \vec{\bf p} }{2|\vec{\bf p} |(2\pi)^3}
 =  |j(p)|^2   \frac{d_R \; d^3 \vec{\bf p}}{2|\vec{\bf p} |(2\pi)^3}
\;\; . 
\eeq{n0}

In the presence of nuclear matter the multiple interactions 
of a fast projectile propagating through  the  medium are modeled 
by a scattering potential assumed to be of 
the form 
\beq V_n = v(q_n)e^{iq_n x_n} \; T_{a_n}(R)\otimes
T_{a_n}(n) = 2\pi \delta(q^0) v(\vec{\bf
q}_n) e^{-i \vec{\bf q}_n\cdot\vec{\bf x}_n} \; T_{a_n}(R)\otimes
T_{a_n}(n) \;\; ,
\eeq{gwmod}
where $T_{a_n}(R)$ is the generator of $SU(N_c)$ in the $d_R$ 
dimensional representation of the jet, and   $T_{a_n}(n)$ is the
corresponding generator in the $d_n$ dimensional representation of
the target. The scattering center is localized at position 
$\vec{\bf x}_n$,  and $v(\vec{\bf q}_n)$ is the Fourier transform of 
the spatial part of the potential. 
We here use $\vec{\bf p}$ to denote the 3D spatial part of a vector $p$ 
and ${\bf p}$ represents its 2D transverse component. 
While there are no {\em a priori} 
limitations  on the functional form of the scattering potential and the 
corresponding differential elastic scattering cross section 
\beq
\frac{d\sigma_{el}(R,T)}{d^2{\bf q}}=
\frac{C_RC_2(T)}{d_A}\frac{|v({\bf q})|^2}{(2\pi)^2}
\;\; , \eeq{sigel}
we here consider 
potentials with a range $1/\mu \ll \lambda$, the mean free path of the 
jet in the medium. This condition together with the 
hard jet approximation ensures the path ordering
of the successive scatterings along the jet trajectory. Diagrams with 
crossed momentum transfers (at points typically separated by 
$z_n-z_{n-1} \simeq \lambda $) are suppressed by 
factors $\sim \exp(-\lambda\mu)$. 
One particular example  used in 
Refs.~\cite{Wang:xy,Gyulassy:2000fs} is the Yukawa 
color-screened potential
\beq
v(\vec{\bf q}_n)\equiv
\frac{4\pi \alpha_s}{\vec{\bf q}_n^{\;2}+\mu^2}=
\frac{4\pi \alpha_s}{(q_{nz}+i\mu_{n})(q_{nz}-i\mu_{n})}
\;\; ,
\eeq{vq}
where $\mu_{n}^2=\mu_{n\perp}^2 = \mu^2+{\bf q}_{n}^{\;2}$ and
$\lambda \mu \gg 1$.

This paper is organized as follows: in Sec.~II  we  present a systematic
recursive way of keeping track of the multiple collision  Feynman  
diagrams in 
terms of ``direct" and ``virtual" interactions. 
We discuss the color and kinematic structure of the single Born (direct) 
operator $\htD$ and the double Born (virtual) operator $\htV$.
In Sec.~III the corresponding  elastic Reaction Operator 
$\htR=\htD^\dagger \htD  + \htV^\dagger + \htV$ is computed. We find 
a recursion relation for the observed inclusive jet distribution  at any 
fixed order in opacity and use impact parameter space resummation to recover
a classical Glauber-like formula. In Sec.~IV we discuss the implications
of our results in relation to the frequently employed Gaussian 
approximation of transverse momentum distribution of jets. We show, however,
that for $\chi<10$  the Gaussian
approximation fails to reproduce the power law tails of the jet $p_T$ 
distribution as well as the $\sim \log Q_{max}^2$  enhancement
of $\left\langle {\bf p}_{T}^2 \right\rangle$.

\section{Tensorial book-keeping and diagrammatic calculus}

The interaction Hamiltonian in a medium with $N$ scattering centers 
in the high jet energy  spinless approximation is given by
\beqar
H_I(t)&=&\int d^3 \vec{\bf x} \, \sum_{i=1}^N
v(\vec{\bf x}-\vec{\bf x}_1) T_a(i)
\phi^\dagger(\vec{\bf x},t)T_a(R) \hat{D}(t)\phi(\vec{\bf x},t)
\;\;,
\eeqar{hj}
where $\hat{D}(t)=i\stackrel{\leftrightarrow}{\partial_t}$.
Each interaction of the colored parton propagating through the nuclear
matter results in one power of the scattering
potential Eq.~(\ref{gwmod}) in the diagrammatic expansion. There is no 
limit on the number of scatterings the jet can have  at a  particular 
scattering center at  position $z_n$. Using the terminology of 
Ref.~\cite{Gyulassy:2000fs} we call such interactions single Born, 
double Born, e.t.c.  It can be seen from Eq.~(\ref{sigel}) that two 
momentum transfers resulting from two interactions with the 
scattering potential are needed to produce one power of the 
elastic scattering cross section $\sigma_{el}$.        
In performing the average over the transverse (relative to the 
jet trajectory) position of the scattering center  
$\left\langle \cdots \right\rangle_{A_\perp} = \int d^2 {\bf b}/ A_\perp 
(\; \cdots \; )$ one  order in the optical thickness  (opacity) 
$N \sigma_{el} / A_\perp = L/\lambda$ is 
thus generated in three ways: a.) single Born or ``direct" interaction in the
amplitude and the complex conjugate amplitude; b.) double Born or 
``virtual" interaction in the amplitude  and
no interaction in the complex conjugate amplitude; c.) no interaction in the
amplitude and double Born interaction in the complementary complex
conjugate amplitude.
 More momentum 
transfers at a fixed longitudinal position $z_n$, e.g. a triple
Born interaction, will produce ${\cal{O}}(\alpha^2_s)$ corrections to the 
opacity expansion and are here neglected.

It is therefore sufficient to consider single and double Born terms 
in the opacity expansion approach, i.e. when the jet propagates by a 
scattering center at position $z_n$ it can either miss the center, 
interact once, or interact twice as illustrated in Fig.~1.     
\begin{center}
\vspace*{8.5cm}
\includegraphics{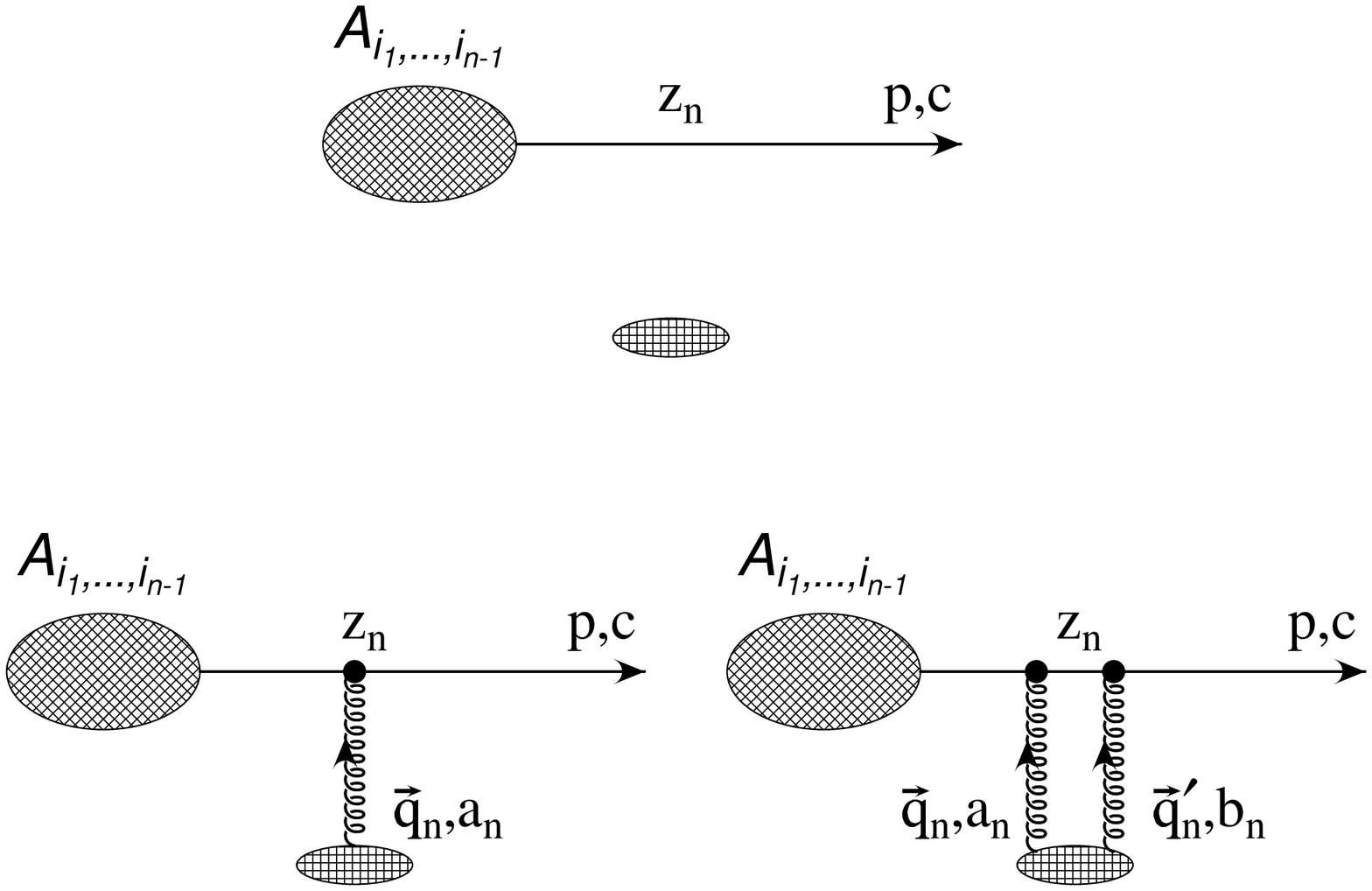}  
\vskip -20pt
\begin{minipage}[t]{15.0cm}
{\small {FIG~1.}
The diagrammatic representation of the interactions with  a fixed  
scattering center located at $z_n$ up to the double Born 
term ( $\, \htI_n \; \vAim$,
$\htD_n \; \vAim$, and $\htV_n \;\vAim \, $) are shown. }
\end{minipage}
\end{center}
\vskip 4truemm
The kinematic and color structure of the scatterings is contained in
the unit $(\,\htI\,)$, direct $(\,\htD \,)$ and virtual $(\, \htV \,)$ 
operators. The path ordering of the
projectile interactions in the medium in the high energy eikonal regime
facilitates the  introduction  of a convenient tensorial book-keeping notation 
that classifies the diagrams according to the type of scatterings 
that the parton has experienced along its trajectory. At order $n$ 
in the opacity expansion there are $3^n$ possible amplitudes classified 
by a set of $n$ indices $i_1 \cdots i_n$.  Here   $i_m = 0$  indicates 
that the jet misses the $m^{th}$ target, 
$i_m = 1$ means that it scatters once, and   
 $i_m = 2$ indicates that it scatters twice. 
The diagrams in Fig.~1 can be algebraically represented as follows:
\begin{eqnarray} 
{\cal A}_{i_1\cdots i_{n-1},0}(p,c) &\equiv& 
\htI  \vAim(p,c) \;\;, \nonumber \\
{\cal A}_{i_1\cdots i_{n-1},1}(p,c) &\equiv& 
\htD  \vAim(p,c)  \;\;,  \nonumber \\
{\cal A}_{i_1\cdots i_{n-1},2} (p,c) &\equiv&  
\htV  \vAim(p,c)  \;\;.  
\label{basit}
\end{eqnarray}  
The set of amplitude indices in Eq.~(\ref{basit}) thus encodes the complete 
history of the jet interactions. Repeating the basic operator steps in  
Eq.~(\ref{basit})  any amplitude that includes parton scatterings inside 
the medium  can be iteratively derived from the unperturbed jet 
production amplitude
\beq
\vAi(p,c)=\prod_{m=1}^n
\left[ \delta_{0,i_m} + \delta_{1,i_m} \htD_m + \delta_{2,i_m} 
\htV_m
\right] M_0(p,c)
\; \; . \eeq{atens}
Time (or longitudinal coordinate)  ordering is implicit in the above formula
for the high energy eikonal limit under consideration here.

To proceed  we need to know the color an kinematic structure of the
direct and virtual interactions. We use  the notation 
$\Delta(p)\equiv (p^2+i\epsilon)^{-1}$ and $\Gamma(p)=p^0$ for the momentum
space representation of the propagator and the vertex factor. 
The vertex factor is treated here in the high  energy spinless limit. 
The direct  iteration step (with $z_n>z_{n-1}> ...$)  
in Eq.~(\ref{basit}) reads
\beqar
{\cal A}_{i_1\cdots i_{n-1},1}(p,c)  &=& 
\int \frac{d^4 q_n}{(2\pi)^4} \; {\cal A}_{i_1 \cdots i_{n-1} }(p-q_n,c) 
 \Delta(p-q_n)v(q_n)\Gamma(2p-q_n) \;
 e^{iq_n(x_n-x_{0})}\, T_{a_n}(n)T_{a_n}(R)\;\;,
\eeqar{p1}
where we have taken into account that the amplitude of order $n-1$
has to be evaluated at momentum $p-q_n$ if the jets emerges on shell 
with momentum  $p$. The double Born amplitude at the {\em same} 
external potential is similarly given by
\beqar
{\cal A}_{i_1\cdots i_{n-1},2}(p,c) &=&
\int \frac{d^4 q_n}{(2\pi)^4}\frac{d^4 q_n^\prime}{(2\pi)^4} 
\; {\cal A}_{i_1 \cdots i_{n-1} }(p-q_n-q_n^\prime,c) 
\Delta(p-q_n-q_n^\prime) \Gamma(2p-2q_n^\prime-q_n) v(q_n) \, 
\times \nonumber \\[.5ex]
&\;& \hspace{.5in}  \times \, \Delta(p-q_n^\prime)\Gamma(2p-q_n^\prime) 
v(q_n^\prime) \;
 e^{i(q_n+q_n^\prime)(x_n-x_{0})} \, T_{a_n}(R)T_{b_n}(R)  
T_{a_n}(n)T_{b_n}(n) \;\; . 
\eeqar{p2}

\section{Elastic Recursion Operator}

The observed jet inclusive distribution 
$dN(p,c) = \sum_{n=0}^\infty dN^{(n)}(p,c)$ 
is expanded in the opacity series, where the  contribution to order
$\chi^n$ is given by
\beq
dN^{(n)}(p,c) = C^N_{n}\; \vAbi(p,c)\vAi(p,c) \equiv
C^N_{n}\; \tr \sum_{i_1=0}^2\cdots \sum_{i_n=0}^2
\bar{\cal A}^{i_1\cdots i_n}(p,c)    
A_{i_1\cdots i_n}(p,c) 
\;\; .\eeq{pnten}
The amplitudes $\bar{\cal A}^{i_1\cdots i_n}(p,c)$  
in Eq.~(\ref{pnten}) are the  complementary  amplitudes given by 
\beqar
\vAbi(p,c)&\equiv& M_0^\dagger
(p,c) \prod_{m=1}^n
\left[ \delta_{0,i_m} \hat{V}_m^\dagger + 
\delta_{1,i_m} \hat{D}_m^\dagger 
+ \delta_{2,i_m} 
\right] 
\; \; .\eeqar{atens2}
The trace is taken over the color matrices and the
binomial coefficient  $C^N_{n}=N!/n!(N-n)! \approx N^n/n!$ is 
introduced to take into account the number of combinations of  
$n$ scattering sites  out of the total $N$.

Performing the sum over the first $n-1$ interaction points and using
Eqs.~(\ref{atens},\ref{atens2}) we obtain a simple recursion identity 
which relates 
$dN^{(n)}$ to $dN^{(n-1)}$ through the Reaction Operator $\htR$ 
\beq
dN^{(n)} = C^N_{n} \;\bar{\cal A}^{i_1\cdots i_{n-1}}
\htR_n {\cal A}_{i_1\cdots i_{n-1}}
\;, \qquad \htR_n = \hat{D}_n^\dagger
\hat{D}_n+\hat{V}_n+\hat{V}_n^\dagger
\; \; . \eeq{rop}

Consider first the direct part  
$dN^{(n)}({\rm Dir.}) \propto \vAbim  \htD^\dagger \htD \vAim$ of the 
Reaction Operator. Performing the color traces we
take into account the identity 
$\tr \, T_a(i)T_b(j)=\delta_{ij}\delta_{ab} C_2(T) d_T/d_A$, which 
enforces that the positions of the ordered scattering centers
in the amplitude and its complementary are identical. We use the form of the 
scattering potential  specified by Eq.~(\ref{gwmod}) as well as the 
the high energy eikonal approximation where the deviations due to the 
in-medium interactions from the initial jet trajectory are small and 
$E^+ \simeq 2 E$.
\beqar
dN^{(n)}({\rm Dir.})
& = & C^N_{n} \;
\int\frac{dq_{z\,n} d^2 {\bf q_n}}{(2\pi)^3} 
\frac{dq_{z\,n}^\prime d^2 {\bf q_n}^\prime}{(2\pi)^3} 
\; \vAbim(p-q_n^\prime) \vAim(p-q_n) \; \frac{C_RC_2(T)}{d_A} \;  
v({\bf \vq}_n)v^*({\bf \vq}_n^\prime)  \, \times \nonumber \\[1ex]
&\;& \times \; 
\frac{E^+}{ E^+q_{z\,n} - q_{z\,n}^2 - {\bf q}_n^2 +i\epsilon} \; 
\frac{E^+}{ E^+q_{z\,n}^\prime - q_{z\,n}^{\prime \,2} - 
{\bf q}_n^{\prime \, 2} -i\epsilon} \;
\left\langle  e^{-i({\bf \vq}_n-{\bf \vq}_n^\prime)\cdot 
({\bf \vx}_n-{\bf \vx}_0)} \right\rangle_{A_\perp} \;\; .
\eeqar{dn1}
The $q_{z\,n}, q_{z\,n}^\prime$ integral can be performed by closing the
contour in the lower/upper half plane. We note that due to the short range
of the scattering potential relative to the mean free path the residues
from the $q_{z\,n} = -i\sqrt{{\bf q}_n^2 + \mu^2}$ and 
$q_{z\,n}^\prime = +i\sqrt{{\bf q}_n^{\prime\,2} + \mu^2}$  poles 
are exponentially suppressed and  only the
residues at  $q_{z\,n} \approx -i\epsilon + {\bf q}_n^2/E^+$ and 
$q_{z\,n}^\prime \approx + i \epsilon + {\bf q}_n^{\prime \, 2}/E^+$ 
contribute. In the high energy limit $E^+\gg \mu \gg {\bf q}_n^2/E^+$ 
the quadratic terms from the residues in the exponent can be neglected.  
This removes the correlations between the scattering centers and there are
no coherence effects at the jet level. We note that this is an important
difference from the case of induced gluon radiation discussed in 
Ref.~\cite{Gyulassy:2000fs}. Emitted gluons are much softer  than the
parent parton and non-negligible $\phi \sim 
({\bf x}_j - {\bf x}_0 )\cdot ({\bf k}-{\bf Q})^2/k^+ $ 
phases control the non-abelian analog  of the Landau-Pomeranchuk-Migdal 
effect~\cite{lpm}.   
Taking into account that  the Glauber thickness function  
(at a fixed impact parameter ${\bf b_0}$)
$T({\bf b_0}) =  \int dz \, \rho({\bf b_0},z) = N/A_\perp$
is expected to vary slowly with impact parameter:
\beq
\left \langle  
e^{-i({\bf q}_n-{\bf q}_n^\prime)\cdot({\bf b} - {\bf b}_0)} 
\right \rangle_{A_\perp} \simeq 
\frac{T({\bf b}_0)}{N} (2\pi)^2 \delta^2({\bf q}_n-{\bf q}_n^\prime) \;\;.
\eeq{expav}  
Eq.~(\ref{expav}) is a key simplifying relation valid in the case
$\sqrt{A_\perp}\gg 1/\mu$ because it
diagonalizes Eq.~(\ref{dn1}) in the transverse momentum variables.

This leads to a simple recursion relation for  the  Direct contribution
to the jet spectrum to the momentum shifted distribution of order $n-1$. 
\beqar
dN^{(n)}({\rm Dir.})
& = & C^N_{n} \;
\int d^2 {\bf q_n} 
\; \vAbim(p,c) \, \left[ \frac{d\sigma_{el}(R,T)}{d^2{\bf q}_n} \,
\frac{T({\bf b}_0)}{N} \; e^{-i {\bf q}_n \cdot {\bf \hat{b}}^\dagger}
\otimes e^{i {\bf q}_n \cdot {\bf \hat{b}} } \right] \,  
\vAim(p,c) \;\; ,  
\eeqar{dit}
where   ${\bf \hat{b}} = i \stackrel{\rightarrow}{\nabla}_{\bf p}$ 
is the  impact parameter operator conjugate to the transverse momentum
${\bf p}$ 
acting to the right, and 
$ {\bf \hat{b}}^\dagger = 
- i \stackrel{\leftarrow}{\nabla}_{\bf p} $ its Hermitian conjugate acting
to the
left.  

Next  consider the virtual contribution
$dN^{(n)}({\rm Vir.}) =  C^N_{n}\; \vAbim  (\htV  + \htV^\dagger ) 
\vAim$ of the  Reaction Operator, Eq.~(\ref{rop}),
\beqar
dN^{(n)}({\rm Vir.})
& = &   C^N_{n}\; 2{\rm Re} \,
\int\frac{dq_{z\,n} d^2 {\bf q_n}}{(2\pi)^3} 
\frac{dq_{z\,n}^\prime d^2 {\bf q_n}^\prime}{(2\pi)^3} 
\; \vAbim(p) \vAim(p-q_n-q_n^\prime) \;  
v({\bf \vq}_n)v({\bf \vq}_n^\prime)  
\; \frac{C_RC_2(T)}{d_A} \, \times \nonumber \\[1ex]
&\;& \times \; \frac{E^+}{ E^+(q_{z\,n}+q_{z\,n}^\prime )- 
(q_{z\,n}+q_{z\,n}^\prime)^2 - ({\bf q}_n+{\bf q}_n^\prime)^2 +i\epsilon} \; 
\frac{E^+}{ E^+q_{z\,n}^\prime - q_{z\,n}^{\prime \,2} - 
{\bf q}_n^{\prime \, 2} +i\epsilon} \, \times \nonumber \\[1ex]
 && \hspace{8cm} \times \,  \left \langle  
 e^{-i({\bf \vq}_n+{\bf \vq}_n^\prime) \cdot (\vec{\bf x}_n-\vec{\bf x}_0)} 
\right \rangle_{A_\perp} \;\; .
\eeqar{dv1}
One notes that the momentum space integrations in this case 
could have  been  performed already at the 
amplitude level, Eq.~(\ref{p2}). Using the same set of 
approximations as for the
Direct part we first perform the $q_{z\,n}$ integral as well as
the impact parameter average and note that the latter constrains
${\bf q}_n + {\bf q}_n^\prime = 0 $.
Picking the residue at $q_{z\,n} \approx - q_{z\,n}^\prime - 
i \epsilon $:
\beqar
dN^{(n)}({\rm Vir.})
& = &  C^N_{n} \;  2 {\rm Re} \,
\int 
\frac{dq_{z\,n}^\prime d^2 {\bf q_n}^\prime}{(2\pi)^3} 
\; \vAbim(p) \vAim(p) \;  
(-i) \frac{T({\bf b}_0)}{N} 
\; \frac{C_RC_2(T)}{d_A} \, \times \nonumber \\[1ex]
&\;& \hspace{6cm}\times \; 
v({\bf \vq}_n^\prime)v(-{\bf \vq}_n^\prime)  
\frac{E^+}{ E^+q_{z\,n}^\prime - q_{z\,n}^{\prime \,2} - 
{\bf q}_n^{\prime \, 2} +i\epsilon} \;\;.
\eeqar{dv2}
It was shown in Ref.~\cite{Gyulassy:2000fs} that for a broad class
of real spherically symmetric potentials of short  range $\ll \lambda$
the remaining $q_{z\,n}^\prime$ integral can be performed. 
The screened  Yukawa potential, Eq.~(\ref{gwmod}), belongs 
in this category when $\mu\lambda\gg 1$. We note that the
contour integration produces a real result and a factor of $\half$ that cancels
the factor of $2$ arising from the Virtual terms in both the amplitude and its 
complementary. The recursion relation for the virtual part of the 
Reaction Operator can then be written as
\beqar
dN^{(n)}({\rm Vir.})
& = &  C^N_{n} \;
\int d^2 {\bf q_n} 
\; \vAbim(p,c) \, \left[ \frac{d\sigma_{el}(R,T)}{d^2{\bf q}_n} \,
\frac{T({\bf b}_0)}{N} \;   ( -1 ) 
\right] \,   \vAim(p,c) \;\; .  
\eeqar{vit}
Combining Eqs.~(\ref{dit},\ref{vit}) one recovers the full structure of the
{\em elastic} Reaction Operator
\beqar
dN^{(n)}(p,c) 
& = &  C^N_{n}
\int d^2 {\bf q_n} 
\; \vAbim(p,c) \, \left[ \frac{d\sigma_{el}(R,T)}{d^2{\bf q}_n} \,
\frac{T({\bf b}_0)}{N} \; \left(  e^{-i {\bf q}_n \cdot {\bf \hat{b}}^\dagger}
\otimes e^{i {\bf q}_n \cdot {\bf \hat{b}} } - 1  \right)\; \right] \,  
\vAim(p,c)  \nonumber \\[.5ex]  
& = & 
\frac{1}{n!} \int \prod_{i=1}^n d^2 {\bf q}_i   
 \left[ T({\bf b}_0) \frac{d\sigma_{el}(R,T)}{d^2{\bf q}_i}
\left(   e^{-{\bf q}_i \cdot \stackrel{\rightarrow}{\nabla}_{\bf p} }
 - 1  \right) \, \right]  \;  dN^{(0)}(p,c) \;\;,
\eeqar{ropit}
where we used the $C^N_{n}/N^n\approx 1/n!$ (Poisson) approximation.

Eq.~(\ref{ropit}) is the main  result of this paper and has transparent
physical interpretation illustrated in Fig.~2. The first term in the kinematic
structure of the  Reaction Operator corresponds to the direct  elastic
scattering part. The next two terms have  delta function strength 
in the forward
direction and  correspond to the virtual corrections (in the amplitude 
and its complementary complex conjugate). These terms enforce
unitarity in the GLV Reaction Operator formalism. Given any initial parton
distribution one can recursively solve from  Eq.~(\ref{ropit}) for the final 
inclusive distribution of jet that have 
penetrated the medium characterized here
by its Glauber thickness function $T({\bf b}_0)$. The method used here is 
quite general and has several important advantages. Within the eikonal 
approximation it provides exact answers to problems related to multiple 
elastic and inelastic processes in QCD media. In particular it was applied 
to the case of gluon radiation\footnote{In the case of medium induced 
gluon bremsstrahlung the Reaction Operator $\htR$ retains its general
operator form Eq.~(\ref{rop}) but its color and kinematic structures are 
different since the on-shell $t=\infty$ 
cuts (see Fig.~2) go through both the jet 
and gluon lines that interact with the double Born term.} 
from jets produced in heavy  ion 
reactions~\cite{Gyulassy:2000fs}. For more details on its uses in 
jet tomography  see Refs.~\cite{Vitev:2001zn,Gyulassy:2000gk}. 
Since the solution is known to all orders in opacity,  cases ranging 
from thin media with few scatterings to thick media with formally 
very large number of scatterings can be studied.   

\begin{center}
\vspace*{8.6cm}
\includegraphics{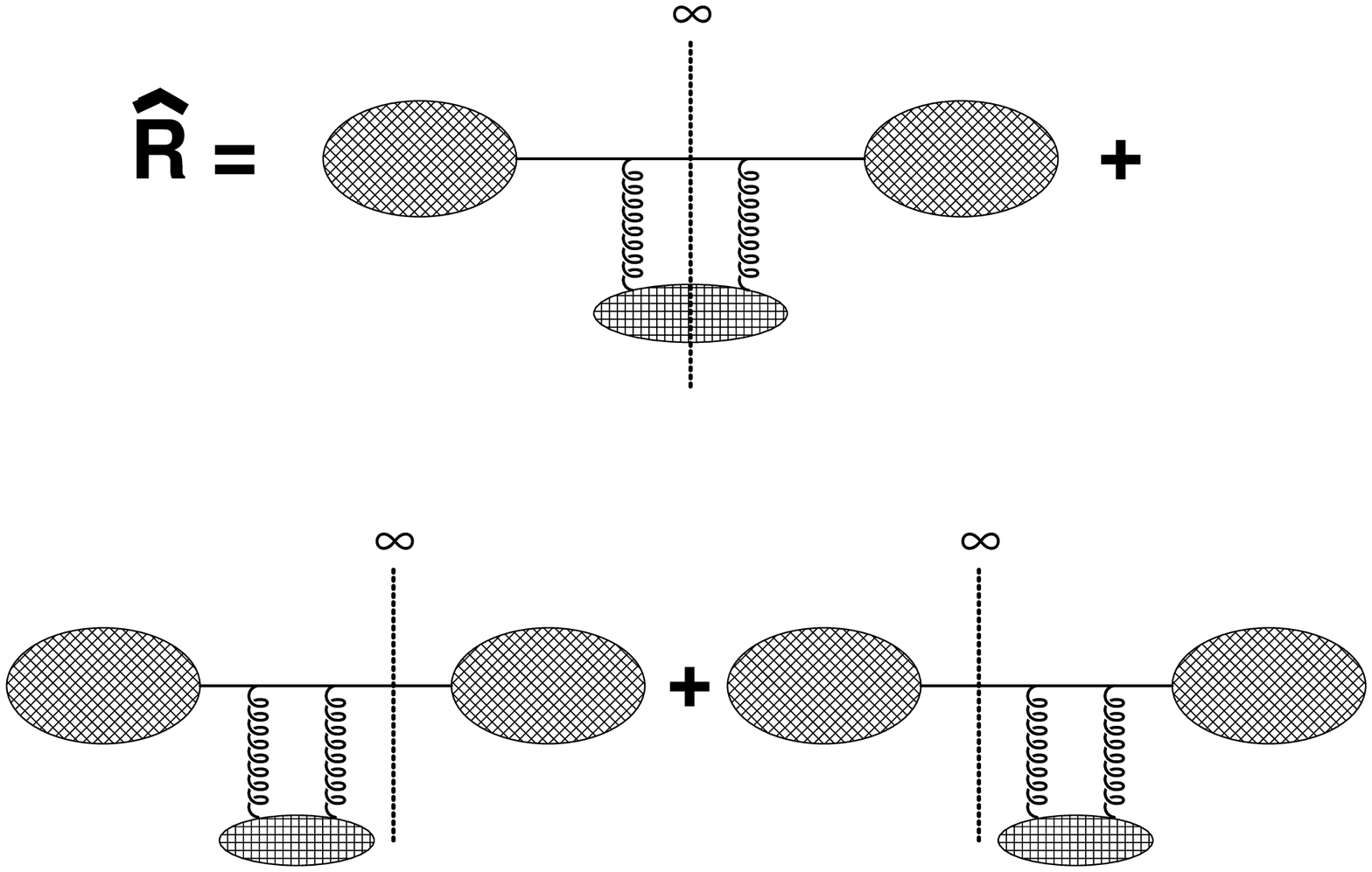}  
\vskip -20pt
\begin{minipage}[t]{15.0cm}
{\small {FIG~2.}
The graphical structure of the Reaction Operator $ \htR = \htD^\dagger \htD 
+ \htV + \htV^\dagger $ is illustrated. It represents
all possible ($t=\infty$) on-shell cuts through a new
double  Born insertion. }  
\end{minipage}
\end{center}
\vskip 4truemm

In the case of multiple elastic scattering,
the reaction operator is so simple that
it is possible to sum over all orders in closed form.
As usual it is most convenient  to perform the summation in impact
parameter space. The distribution in impact parameter space conjugate
to the transverse momentum ${\bf p}$ is
\begin{eqnarray} 
dN({\bf b}) & = & \sum_{n=0}^\infty \frac{1}{n!} 
\left[ T({\bf b}_0) \, \left(  
\tilde{\sigma}_{el}({\bf b}) - 
\sigma_{el}  \right)
\right]^n dN^{(0)}({\bf b}) = 
e^{  T({\bf b}_0) \, \left( 
{\tilde{\sigma}_{el}}
({\bf b})  -  \sigma_{el}^{tot} \right)  }
dN^{(0)}({\bf b})  \;\;,
\label{iptr}
\end{eqnarray} 
where the Fourier transform of the differential cross section  was denoted by 
$(2\pi)^2 d \tilde{\sigma}_{el}/ d^2{\bf q} ({\bf b}) \equiv 
\tilde{\sigma}_{el}({\bf b})$ and $\tilde{\sigma}_{el} ({\bf 0}) = 
\sigma_{el}^{tot}$.
The impact parameter space representation Eq.~(\ref{iptr}) depends only 
on the  difference  $ \tilde{\sigma}_{el}({\bf b}) 
  - \tilde{\sigma}_{el}({\bf 0}) $.
Transforming back to momentum space, we recover
the parton version  of the Glauber multiple collision 
series~\cite{Glauber:1970jm}
\beqar
dN({\bf p}) & = & e^{-\sigma_{el} T({\bf b}_0) } \int d^2 {\bf b} \;
e^{i{\bf p} \cdot {\bf b} }    \,  e^{ 
{\tilde{\sigma}_{el}}
({\bf b}) T({\bf b}_0) } 
dN^{(0)}({\bf b}) \nonumber \\[.5ex]
& = & \sum_{n=0}^\infty e^{-\chi} \frac{\chi^n}{n!}  
\int \prod_{i=1}^n d^2 {\bf q}_i \frac{1}{\sigma_{el}} 
 \frac{d \sigma_{el} }{d^2 {\bf q}_i}  \;  
dN^{(0)}({\bf p} -{\bf q}_1 - \cdots - {\bf q}_n)  \;\;, 
\label{glau}
\eeqar{glaberf}
 where $\chi =\sigma_{el} T({\bf b}_0)
=L/\lambda $. Eq.~(\ref{glau}) has the familiar physical interpretation
of a Poisson random walk in  transverse momentum space  distributed
according to  $d \sigma_{el}/\sigma_{el}$.  

\section{Discussion}

Impact parameter space resummation is often used in conjunction with 
small impact parameter approximation of the differential cross section,
Eq.~(\ref{sigel}). We show below that although such an approach is  
analytically appealing, it gives jet distributions that may differ
substantially from the exact formula. 

The Fourier transform of the {\em normalized} 
cross section is given by:
\beq
\frac{d \tilde{\sigma}_{el} }{d^2 {\bf q}}({\bf b}) = 
\int \frac{d^2 {\bf q}}{(2 \pi)^2} \; e^{-i{\bf q} \cdot {\bf b}}
\frac{1}{\pi} \frac{\mu^2}{({\bf q}^2+\mu^2)^2} = 
\frac{\mu \, b}{4 \pi^2} K_1(\mu\, b)   \approx 
\frac{1}{4 \pi^2} \left(1- \frac{\xi \, \mu^2\, b^2}{2}
+ {\cal O}(b^3)  \right) \;\;,
\eeq{ftdc} 
where  $b=|{\bf b}|$ and in the quadratic term in Eq.~(\ref{ftdc}) 
the $\log 2/(1.08\, \mu\, b) $ multiplicative factor has been absorbed
into a $b$-independent constant $\xi$ due to its  small 
logarithmic variation. We consider the
transverse momentum broadening~\cite{Blaizot:1986ma,Gupta:1994mk}
of a jet propagating in the ``$\hat{z}$'' 
direction, i.e. $dN^{(0)}/d^2 {\bf p} = \delta^2({\bf p})$. In the 
approximation 
that dominantly  small impact parameter scatterings 
are important, the
momentum space distributions from Eqs.~(\ref{iptr},\ref{glau})
reduce to the  classic Moliere form~\cite{Wiedemann:2000za}
\beq
dN({\bf p}) = \int d^2{\bf b} \; e^{i{\bf p}\cdot {\bf b} } 
\frac{1}{(2\pi)^2} \frac{e^{-\frac{\chi\, \mu^2 \, \xi \, b^2}{2}}  }
{\chi\, \mu^2 \, \xi } 
= \frac{1}{2\pi} \frac{e^{-\frac{p^2}{2\, \chi  \, \mu^2 \, \xi} }}
{\chi\, \mu^2\, \xi }
\;\;.
\eeq{gauss}
The resulting distribution is of Gaussian form and has a width of
$\chi\, \mu^2\, \xi$, 
i.e. $\left\langle {\bf p}^2  \right\rangle = \chi\, \mu^2\, \xi$.   

\begin{center}
\vspace*{9.4cm}
\includegraphics{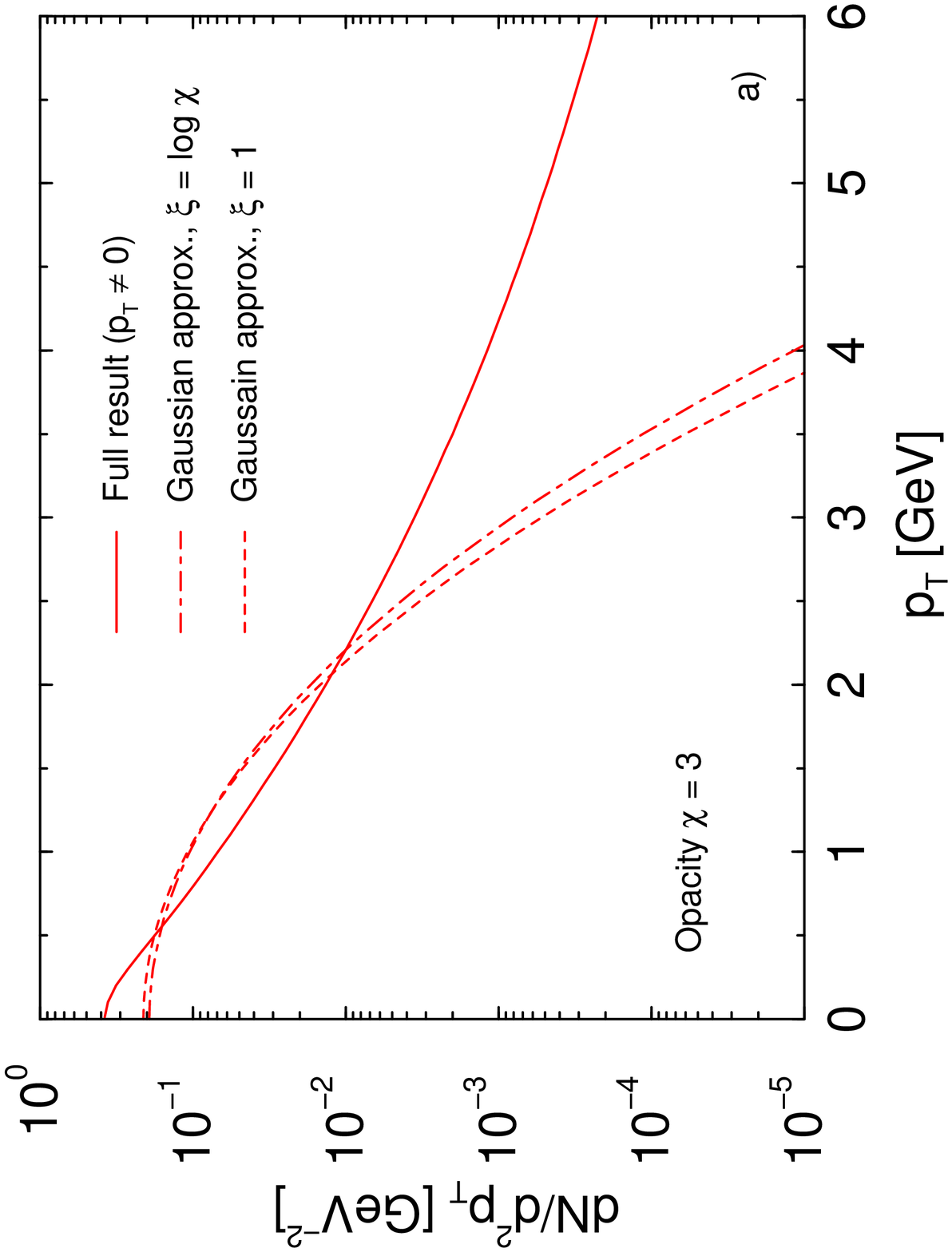}  
\includegraphics{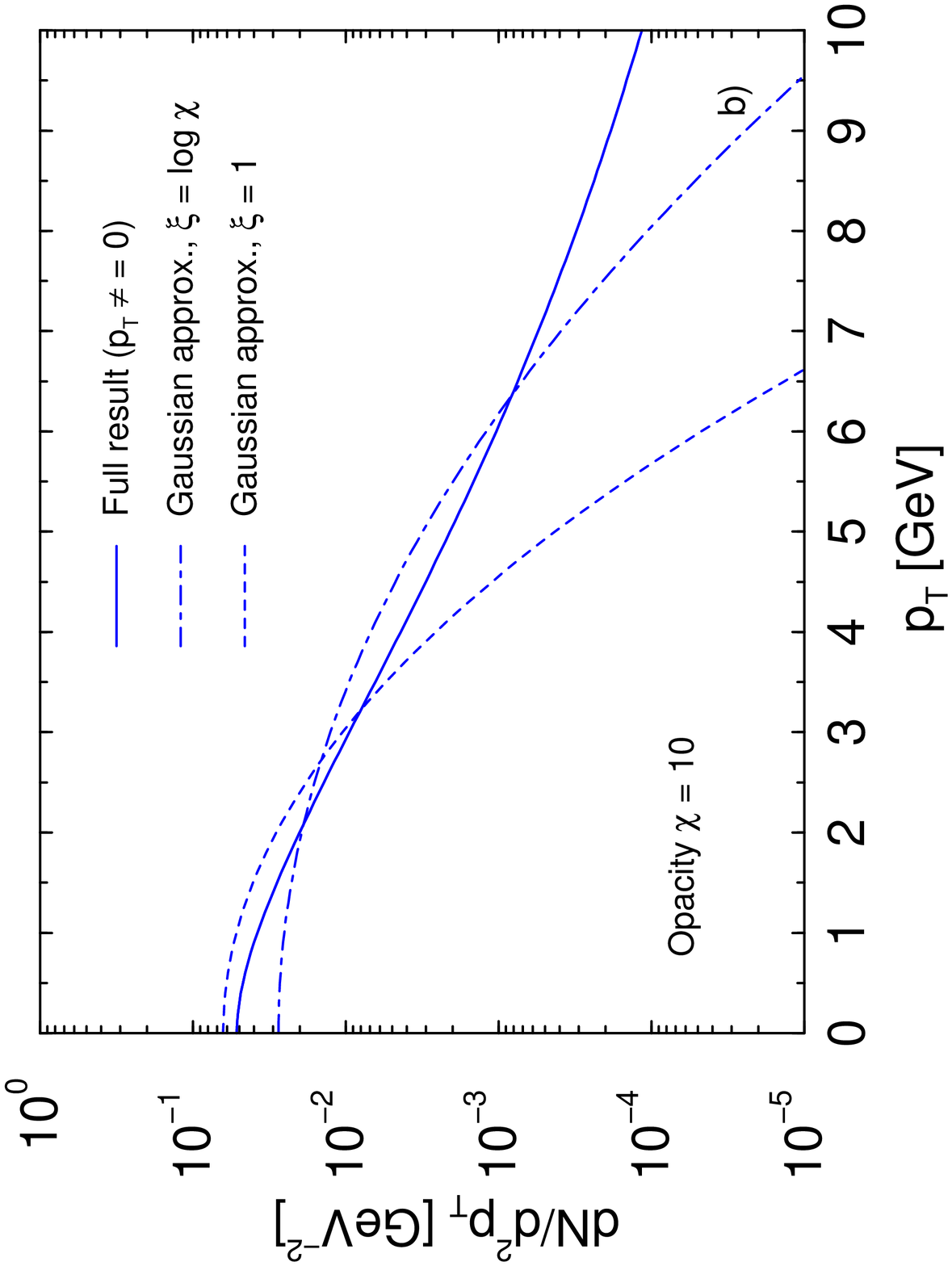}  
\vskip -20pt
\begin{minipage}[t]{15.0cm}
{\small {FIG~3.}
The final jet $p_T$ distribution is shown versus $p_T$ for two different
opacities $\chi =3$ (Fig.~3a) and $\chi =10$ (Fig.~3b). We compare the
full result (without the delta function  contribution at $p_T \sim 0$) to
the Moliere Gaussian approximation with $\xi=1$ and $\xi=\log\chi$. 
In this example we use $\mu^2=0.25$~GeV$^2$. }
\end{minipage}
\end{center}
\vskip 4truemm

To asses the accuracy of the Gaussian ansatz for finite $\chi$
relevant in nuclear physics, we compare the Fourier
transform of Eq.~(\ref{iptr}) to Eq.~(\ref{gauss}). Note that in 
going from Eq.~(\ref{iptr}) back to  momentum  space,  a 
$e^{-\chi}\,\delta^2({\bf p})$  component arises. This component is 
of course not displayed, and for finite $p$ 
numerical integration converges much more
rapidly if unity is subtracted from $\exp[
\tilde{\sigma}({\bf b})T({\bf b_0})]$
in the integrand (\ref{glau}).

In Figs.~3a, 3b numerical results are shown for  
 opacities $\chi = 3$ and $\chi = 10$. 
For large opacities $\chi \geq 10$  (see Fig.~3b) the Gaussian approximation
gives better results than for small ones $\chi \leq 5$. However,
as is well known, at larger
transverse momenta it fails to account for the power law $p_T$ tail of the
inclusive distribution. Unlike familiar Moliere scattering in 
atomic matter, there are no form factors to limit the growth
of the high $p_T$ Rutherford tail out 
to the kinematic limit $Q^2_{max} \sim E_0 \mu $
in the case of QCD.  Our results for the rms transverse momentum
kick shown in Fig. 4  demonstrate that 
in semi-hard and hard pQCD processes there is a logarithmic in $Q$ 
enhancement of the mean squared transverse momentum, 
$\left\langle {\bf p}^2  \right\rangle \simeq 
\chi\, \mu^2 \log (1+Q^2/\chi \mu^2)$ that is missed in the 
dipole approximation. Of course for large $Q$ radiative energy loss
also contributes substantially to the $p_T$ broadening
and should be taken into account in applications.
\begin{center}
\vspace*{9.4cm}
\includegraphics{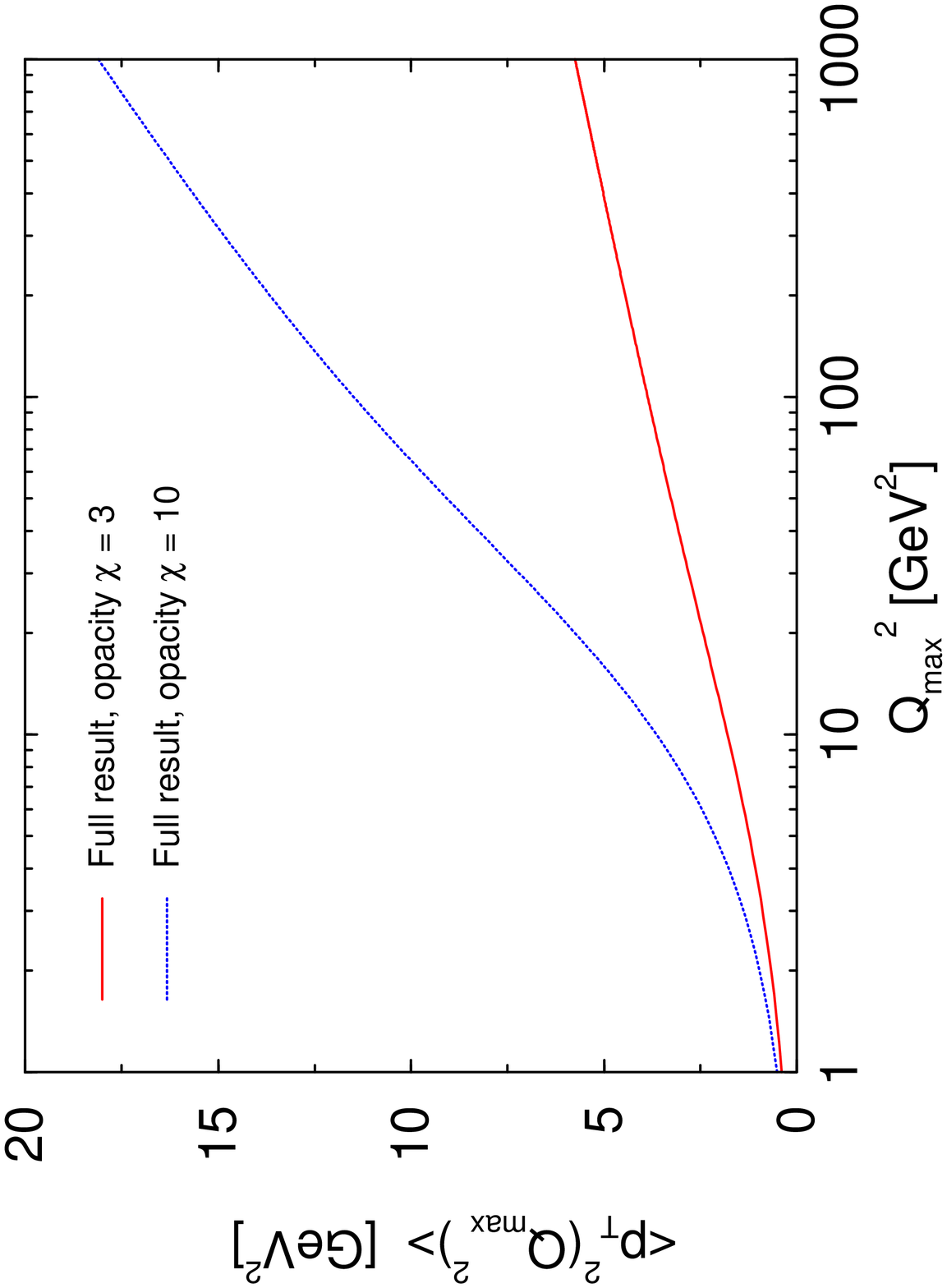}  
\vskip -20pt
\begin{minipage}[t]{15.0cm}
{\small {FIG~4.}
The transverse momentum broadening of jets is shown versus $Q_{max}$ 
for two different opacities $\chi =3, 10$ as in Figs.~3a, 3b. 
We here use $\mu^2=0.25$~GeV$^2$ for illustration.} 
\end{minipage}
\end{center}
\vskip 4truemm

In summary, we showed that the Reaction Operator formalism
developed in~\cite{Gyulassy:2000fs} easily recovers the well
known Glauber limit for elastic multiple collisions in pQCD matter
and sheds light on the accuracy of the dipole approximation. 
In the future it would be interesting to combine both elastic and 
the LPM suppressed inelastic processes in this method and apply
the results to jet acoplanarity observables suggested in
Refs.~\cite{Blaizot:1986ma,Gupta:1994mk}.

\acknowledgments

This work was supported by the Director, Office of Science, 
Office of High Energy and Nuclear Physics,
Division of Nuclear Physics, of the U.S. Department of Energy
under Contract No. DE-FG02-93ER40764 and by the U.S. NSF under INT-0000211 
and  OTKA No. T029158.


\end{document}